\documentclass[11pt,a4paper]{article}
\usepackage{jheppub}
\pdfoutput=1

\usepackage[]{amsmath}
\usepackage[]{amssymb}
\usepackage{amscd}
\usepackage{amsfonts}
 \usepackage{amsthm}
  \usepackage{ytableau}

\newcommand{\I}{\mathrm{i}}
\newcommand{\dd}{\mathrm{d}}
\newcommand{\ii}{\mathrm{i}}
\newcommand{\be}{\begin{equation}}
\newcommand{\ee}{\end{equation}}

\def\XXint#1#2#3{{\setbox0=\hbox{$#1{#2#3}{\int}$}
     \vcenter{\hbox{$#2#3$}}\kern-.52\wd0}}

\usepackage{hyperref}

\usepackage[toc,page]{appendix}

\usepackage[utf8]{inputenc}
\usepackage[english]{babel}

 \usepackage{graphicx}
 
 \usepackage{enumerate}

\title{Multiple phases in a generalized Gross-Witten-Wadia matrix model}
\author{Jorge G. Russo $^\dagger$}
\author{and Miguel Tierz $^\ddagger$ $^{\dagger\dagger}$}

\affiliation{$\dagger$ {\it Instituci\'o Catalana de Recerca i Estudis Avan\c{c}ats (ICREA), }
\\
{\it Pg. Lluis Companys, 23, 08010 Barcelona, Spain.}
\\
{\it  Departament de F\' \i sica Cu\' antica i Astrof\'\i sica and Institut de Ci\`encies del Cosmos,}
\\
{\it Universitat de Barcelona, Mart\'i Franqu\`es, 1, 08028 Barcelona, Spain. }
}
\affiliation{$\ddagger$ Departamento de Matem\'{a}tica, Faculdade de Ci\^{e}ncias, ISCTE - Instituto Universit\'{a}rio de Lisboa, Avenida das For\c{c}as Armadas, 1649-026 Lisboa, Portugal.}
\affiliation{$\dagger\dagger$ Grupo de F\'{i}sica Matem\'{a}tica, Departamento de Matem\'{a}tica, Faculdade de Ci\^{e}ncias, Universidade de Lisboa, Campo Grande, Edif\'{i}cio C6, 1749-016 Lisboa, Portugal.}
\emailAdd{jorge.russo@icrea.cat}
\emailAdd{mtpaz@iscte-iul.pt}
\emailAdd{tierz@fc.ul.pt}

\abstract{We study a unitary matrix model of the Gross-Witten-Wadia type, extended with the addition of characteristic polynomial insertions. The model interpolates between solvable unitary matrix models and is the unitary counterpart of a deformed Cauchy ensemble. Exact formulas for the partition function and Wilson loops are given in terms of Toeplitz determinants and minors and large $N$ results are obtained by using Szegö theorem with a Fisher-Hartwig singularity. 
In the large $N$ (planar) limit with two scaled couplings,
the theory exhibits a surprisingly intricate phase structure in the two-dimensional parameter space.
}

\begin{document}

\maketitle
	
\section{Introduction}

	The study of random matrix ensembles has been a subject developing for many decades already. The array of applications is vast by now and ever expanding \cite{Mehta,Forrester,BDS}. Among the many aspects of random matrices that can be studied is the analysis of phase transitions in a so-called double scaling limit. 
	
	This type of analysis is relevant for example in the study of phase transitions in gauge theories. When combined with localization results, which effectively provide integral representation, of random-matrix type, of observables of the gauge theory, such as partition functions or Wilson loop averages, this analysis is especially potent and has led to many novel results in the study of supersymmetric and topological gauge theories.

In this context, the matrix models are typically and predominantly models of Hermitian random matrices. However, another important type of matrix models are the unitary matrix models, such as Dyson's circular ensembles \cite{Mehta} or the matrix model of Gross and Witten \cite{Gross:1980he} and Wadia \cite{Wadia:2012fr,Wadia:1980cp}, for example. 

The unitary group $U(N)$ with Haar measure   
has eigenvalue probability density function (see \cite[Chapter 2]{Forrester}, for example)
\begin{equation}\label{unit}
   \frac{1}{(2 \pi )^N N!} \prod_{1 \le j < k \le N} | z_k - z_j |^2,
 \quad z_l := e^{i \theta_l} \in \mathbb{T}, \quad \theta_l \in (-\pi,\pi] ,
\end{equation} 
where $\mathbb{T} = \{z \in \mathbb{C}: |z|=1 \} $.
By unitary matrix model, we refer to the setting where one studies averages over $U \in U(N)$ of functions $ w(U) $, which are symmetric functions of the eigenvalues of $ U $ only (class functions) with 
the factorization property $ \prod_{l=1}^N w(z_l) $ for 
$ \{z_1,\dots, z_N \} \in {\rm Spec}(U) $. 
Writing down the Fourier components $\{w_l\}_{l\in \mathbb{Z}}$ of the weight
$ w(z) = \sum_{l=-\infty}^\infty w_l z^l $, it holds that
\cite{Forrester}
\begin{equation}\label{Toepl}
  \Big \langle \prod_{l=1}^N w(z_l) \Big \rangle_{U(N)} =
  \det[ w_{i-j} ]_{i,j=1,\dots,N}, 
\end{equation}
relating the partition function with the determinant of a Toeplitz matrix, whose entries are the Fourier coefficients of the weight function. We will make use of this equivalent Toeplitz determinant formulation below.

As we shall explain in the next Section, the Gross-Witten-Wadia (GWW) model corresponds to the case where the weight function is:
\be
\omega (z)=\exp (t(z+z^{-1})),
\ee
where $t$ is a real parameter, although we will be using the original physics notation in \cite{Gross:1980he}, in terms of a coupling constant.

One of the main aspects associated with this model is that it exhibits a third-order phase transition at large $N$ \cite{Gross:1980he,Wadia:2012fr}. The study of this transition has lead to many delicate results and its detailed analysis also encompasses many influential works in mathematics as well, involving also the Tracy-Widom law and culminating in the seminal solution of the long-standing conjecture (as it was proposed by Ulam in the early 1960s) on the longest increasing subsequence of a random permutation, by Baik, Deift and Johansson \cite{BDJ,BDS,Romik}.

On  the other hand, models involving other weights, that correspond to a pure Fisher-Hartwig (FH) singularity in the context of Toeplitz matrices \cite{Toep-book}, such as 
\be
\omega \left( z\right) =(1+z)^{\alpha }(1+z^{-1})^{\beta }\ ,
\label{mod2}
\ee
are actually solvable even at finite $N$ and have been also thoroughly studied, with many mathematical applications in problems in combinatorics, representation theory and number theory \cite{Forrester,BDS}. Here $\alpha$ and $\beta$ are real parameters that will be equal in our discussion below.
Both models emerge in the study of probability measures associated to partitions (see \cite{TW}, for example).

In this work, we shall be studying a unitary matrix model that is made of both terms and we will be stressing its large $N$ behavior and the ensuing phase transitions in scaling limits involving the two parameters present.

\section{New unitary matrix model from deformation of GWW model}

By an appropriate axial gauge choice,  two-dimensional  lattice $U(N)$ gauge theory with Wilson lattice action is effectively described by a random unitary matrix model, with partition function
\be
Z=  \int dU e^{\frac{1}{g^{2}}\mathrm{Tr}(U+U^{\dagger })},
\ee
where $g$ is the coupling constant and only parameter in the theory. Integrating over the volume of the $U(N)$ group following a standard procedure, one arrives at
\begin{equation}
Z_{N}=\int_{(0,2\pi ]^{N}}\prod_{1\leq j<k\leq N}\left\vert e^{\ii\varphi
_{j}}-e^{\ii\varphi _{k}}\right\vert ^{2}\prod_{j=1}^{N} \exp \left( \frac{2}{g^2}\cos \left( \varphi_{j}\right) \right) \frac{\dd\varphi _{j}}{2\pi}.
\label{GWWpar}
\end{equation}%
This is the celebrated  Gross-Witten-Wadia matrix model \cite{Gross:1980he,Wadia:2012fr} alluded above. It has generated a large amount of interest, spanning now four decades. As mentioned above, chief among the reasons for this interest, is the existence of a large $N$ third order phase transition in a double scaling limit. 

In this work, we shall study a generalized version of the model, emphasizing the analysis of its phase structure in a number of scaling limits. To obtain our model, we start  with the Hermitian matrix model introduced in \cite{Russo:2020pnv},
with potential
\be
V(M)= A\, {\rm Tr}\ln (1+M^2)+B\, {\rm Tr}\frac{1}{1+M^2}\ .
\ee
The partition function is given by
\be
Z=\frac{1}{N!}\int \frac{d^Na}{(2\pi)^N}\, \prod_{i<j}^N (a_i-a_j)^2 \, \prod_{i=1}^N 
\frac{\exp[-B/(1+a_i^2)]}{(1+a_i^2)^A}\ .
\label{partis}
\ee
We pass from angular variables to the real line using the usual one-dimensional stereographic projection:

\begin{equation*}
e^{\ii\varphi }=\frac{1+\ii x}{1-\ii x},\qquad 
-\pi < \varphi  < \pi ,\ x\in {\mathbb{R}}.
\end{equation*}%
Thus, we have the following unitary matrix model:%

\begin{equation}
Z_{N}=\frac{e^{-\frac12 BN}}{N!}\int_{(0,2\pi ]^{N}}\prod_{1\leq j<k\leq N}\left\vert e^{\ii\varphi
_{j}}-e^{\ii\varphi _{k}}\right\vert ^{2}\prod_{j=1}^{N}\cos ^{2\nu}\left( 
\frac{\varphi _{j}}{2}\right) \exp \left( -\frac12 B\cos \left( \varphi
_{j}\right) \right) \frac{\dd\varphi _{j}}{2\pi}\ .
\label{moregeneral}
\end{equation}%
with $\nu\equiv A-N$. The shift $A\to  A-N$ takes into account
the Jacobian of the transformation from $a_j$ to $\varphi_j$
(see \cite{WitteForr} for the similar calculation in the Cauchy ensemble).
As explained in the introduction, this matrix model interpolates between two matrix models. On one hand, for $\nu =0$, it is the celebrated and influential Gross-Witten-Wadia model and, on the other hand, for $B=0$ is the exactly solvable matrix model \eqref{mod2}, which is actually very central in random matrix theory \cite{Forrester} and in the theory of Toeplitz banded matrices \cite{Toep-book}.
Indeed, for $B=0$ the matrix model is\footnote{Along the paper, we will be using the different equivalent ways of writing
the term in the matrix model: $\cos ^{2\nu }\left( \frac{\theta }{2}\right) =%
\frac{1}{2^{\nu }}(1+\cos \theta )^{\nu }=\frac{(1+z)^{\nu }(1+z^{-1})^{\nu }%
}{2^{2\nu }}=\frac{\left\vert 1+z\right\vert ^{2\nu }}{2^{2\nu }}.$
}
\begin{equation}\label{model2}
  \frac{1}{2^{2N\nu }}\Big \langle \prod_{l=1}^N \left\vert 1+z_{l}\right\vert ^{2\nu } \Big \rangle_{U(N)}\ , 
\end{equation}
which is the average of a characteristic polynomial over the Circular Unitary Ensemble (CUE) \eqref{unit} \cite{WitteForr,Forrester} and is also a particular case of the pure FH weight \eqref{mod2}, when $\alpha=\beta=\nu$. 
From the point of view of the right hand side of \eqref{Toepl}, that is, as a Toeplitz determinant, it corresponds to the pure Fisher-Hartwig singularity case, and its exact evaluation is known \cite{FH,Toep-book}:
	\begin{equation}
		\label{eq:exactZEt1}
			\mathcal{Z}_{N} (B=0) = \frac{G(N+1)}{2^{2\nu N }}\frac{ G (\nu + 1 )^2 G (2\nu + N +1) }{ G (2\nu + 1) G (\nu + N + 1)^2 }\ ,
		\end{equation}
where $G( \cdot )$ is the Barnes $G$-function, which, for integer values of the argument is $G(n) = \prod_{j=0} ^{n-2} j!$. This result holds for complex $\nu$ with $\Re\left( \nu \right) >-\frac{1}{2}$ \cite{FH}.




\vspace{0.1cm}

In these two limits, the matrix model partition function is a tau function of Painlevé III', in the GWW case, or a tau function of Painlevé V for a generalization of \eqref{model2} \cite{Forrester}. Among many other aspects, the connection with Painlevé is useful for resurgence computations \cite{Ahmed:2017lhl,Ahmed:2018gbt}.

\vspace{0.1cm}

\subsection{Finite $N$ partition function and Wilson loops}

We can adopt the Toeplitz determinant point of view \eqref{Toepl} for finite $N$ evaluations of the partition function. Let us consider the case when $\nu $ is a positive integer. The relationship between our matrix model and the ordinary GWW matrix model is not that of a parametric derivative with regards to $B$. However, the weight function $\omega \left( \varphi \right) $ of our matrix model has this property, as it satisfies
\begin{equation}
\omega \left( \varphi \right) =\frac{1}{2^\nu }\left( 1+\cos \left( \varphi \right) \right)
^{\nu }e^{-\frac12 B(1+\cos \left( \varphi \right) )}=(-1)^{\nu }\frac{d^{\nu
}\omega _{\nu =0}(\varphi )}{dB^{\nu }},
\end{equation}%
for positive integer $\nu .$ It has the well-known expansion

\begin{equation}
\omega _{\nu =0}\left( \varphi \right) =e^{-B/2}\sum_{n\in \mathbb{Z}%
}   (-1)^n\ I_{n}(B/2)e^{in\varphi },
\label{GWexpa}
\end{equation}%
where $I_{n}(B)$ are the modified Bessel functions of first kind. The derivative of these Bessel functions can be written as a recurrence relation\footnote{The second identity can be interpreted as the symmetric random walk equation.}:
\begin{eqnarray}
\frac{dI_{0}(x)}{dx} &=&I_{1}(x),\\
\frac{dI_{n}(x)}{dx} &=&\frac{1}{2}(I_{n-1}(x)+I_{n+1}(x)).
\label{1st}
\end{eqnarray}
Although the prefactor $e^{-NB/2}$ in \eqref{moregeneral} is not essential for the definition of the  model,
as one may define the matrix model without this prefactor, we can take into account
by considering derivatives of $e^{-B/2}I_{n}(B/2)$, not just the derivative of the
Bessel function, otherwise we would be generating the term $\cos \theta $ in
the matrix model, instead of $\cos \left( \theta /2\right) $. 
One can use the formula%
\be
\frac{d^{\nu}(e^{-x}I_{n}(x))}{dx^{\nu}}=\sum\limits_{k=0}^{\nu}(-1)^{\nu-k}\binom{\nu%
}{k}e^{-x}\frac{d^{k}\left( I_{n}(x)\right) }{dx^{k}} \ .
\ee
The $kth$ derivative of the Bessel function admits several interesting
expressions, including hypergeometric ones. The simplest one is:%
\be
\frac{d^{k}\left( I_{n}(x)\right) }{dx^{k}}=2^{-k}\sum\limits_{j=0}^{k}%
\binom{k}{j}I_{2j-k+n}(x).
\ee
We have then that (\ref{moregeneral}) is the determinant:
\begin{equation}
  Z_{N}(\nu,B)=\frac{1}{2^{\nu N}}\det[ w_{i-j} ]_{i,j=1,\dots,N}, 
  \label{detw}
\end{equation}
with 
\begin{equation}
  w_{i-j}=(-1)^{i-j}\sum\limits_{k=0}^{\nu}\sum\limits_{l=0}^{k}2^{-k}\binom{\nu}{k}\binom{k}{l}e^{-\frac{B}{2}}I_{2l-k+i-j}(\frac{B}{2}).
  \label{entries}
\end{equation}
As for particular cases, we see that $\nu=0$ corresponds indeed to GWW model
with the prefactor $e^{-BN/2}$ and, for $\nu=1$, we have that the matrix entry $(i,j)$
is:
\begin{equation}
(-1)^{i-j}e^{-\frac{B}{2}}\left( I_{i-j}(\frac{B}{2})+2^{-1}I_{i-j-1}(\frac{B%
}{2})+2^{-1}I_{i-j+1}(\frac{B}{2})\right) ,
\end{equation}
in consistency with \eqref{1st}. Notice that there is no problem if $\nu>n$ because it holds that $I_{-\alpha }(\frac{B}{2})=I_{\alpha }(\frac{B}{2}).$ Expressions for higher values of $\nu$ and specific values of $N$ can be quickly generated, case-by-case (as happens with the GWW model) with \eqref{entries}.

We note that the same result can be obtained, without having to consider any derivative of Bessel functions, by carrying out the product of the term $\left( 1+\cos \left( \varphi \right) \right)
^{\nu }$, which is a polynomial in the variables $e^{i\varphi }$ and $e^{-i\varphi }$, with \eqref{GWexpa}, and then identifying again all the coefficients corresponding to a given power of $z=e^{i\varphi }$. By using the trinomial theorem (or twice the binomial theorem), it is immediate that:
\begin{equation}
\left( 1+\cos \left( \varphi \right) \right)
^{\nu }=\sum_{k=0}^{\nu }\sum_{l=0}^{k}\frac{1}{2^{k}}\binom{\nu }{k}\binom{k }{l}z^{2l-k},
\end{equation}
where $z=e^{i\varphi }$. The product of this expansion with the Bessel expansion of the remaining term, Eq. \eqref{GWexpa} above, leads then again to \eqref{entries}.

The determinant representation can be applied directly to the study of averages of product of two traces, in arbitrary representations $\tilde r$ and $ r$:
\be
\left\langle W_{N}^{r ,\tilde r }\right\rangle =\left\langle \mathrm{Tr}%
_{r }(U)\mathrm{Tr}_{\tilde r }(U^{\dag })\right\rangle =\frac{1}{%
Z_{N}\left( B,\nu \right) }\int dU\omega(U)s_{r }(U)s_{\tilde r }(U^{\dag }). 
\ee
This describes, when we only have one trace (taking the void partition, $\tilde r=\varnothing$, for example), Wilson loops in arbitrary representation. The determinant description is now:
\be
\left\langle W_{N}^{r ,\tilde r }\right\rangle =\frac{\det[w_{i-\tilde r_{i}-j+r_{j}} ]_{i,j=1,\dots,N}}{\det[ w_{i-j} ]_{i,j=1,\dots,N}}.
\label{av}
\ee
Hence, the un-normalized average is a minor of the matrix in \eqref{detw}. The striking pattern of columns and rows is described in \cite{Minors}, but can be read off from \eqref{av} as well.

Additionally, there is an extension of Szegö theorem, applicable to these averages\footnote{See \cite{Minors} for comments on the validity of the formula when there are Fisher-Hartwig singularities in the weight function.} \cite{Minors}. As with the partition function, the result is expressed in terms of the Fourier coefficients of the potential. We summarize a few cases for the model at hand in the Table 1.

\ytableausetup{boxsize=0.2cm}
\begin{table}
\begin{center}
\begin{tabular}{|c|c|l|c|c|l|}
\hline
$\tilde r$ & $ r$ & $\lim_{N\rightarrow\infty} \left\langle W_{N}^{r ,\tilde r }\right\rangle$ & $\tilde r$ & $r$ & $\lim_{N\rightarrow\infty} \left\langle W_{N}^{r ,\tilde r }\right\rangle$\\
\hline

$\varnothing$ & $\ydiagram{1}$ & $\nu-\frac{B}{4}$ & $\varnothing$ & $\ydiagram{2}$ & $\frac{1}{2}(\nu-\frac{B}{4})^{2}-\frac{\nu}{2}$\\
$\varnothing$ & $\ydiagram{1,1}$ & $\frac{1}{2}(\nu-\frac{B}{4})^{2}+\frac{\nu}{2}$ & $\varnothing$ & $\ydiagram{3}$ & $\frac{1}{6}(\nu-\frac{B}{4})^{3}-(\nu-\frac{B}{4})\frac{\nu}{2}+\frac{\nu}{3}$ \\
$\varnothing$ & $\ydiagram{1,1,1}$ & $\frac{1}{6}(\nu-\frac{B}{4})^{3}+(\nu-\frac{B}{4})\frac{\nu}{2}+\frac{\nu}{3}$ & $\varnothing$ & $\ydiagram{2,2}$ & $\frac{1}{12}(\nu-\frac{B}{4})^{4}-(\nu-\frac{B}{4})\frac{\nu}{3}+\frac{\nu^{2}}{4}$ \\
\hline 
\multicolumn{6}{c}{} \\
\hline 

$\tilde r$ & $r $ & \multicolumn{4}{l|}{$\lim_{N\rightarrow\infty} \left\langle W_{N}^{r ,\tilde r }\right\rangle$}  \\
\hline
$\ydiagram{1,1}$ & $\ydiagram{1,1}$ & \multicolumn{4}{l|}{$\frac{1}{4}(\nu-\frac{B}{4})^{4}+(\nu-\frac{B}{4})^{2}+\frac{\nu}{2}(\nu-\frac{B}{4})^{2}+\frac{\nu^{2}}{4}+1$} \\
$\ydiagram{1}$ & $\ydiagram{3}$ &  \multicolumn{4}{l|}{$\frac{1}{6}(\nu-\frac{B}{4})^{4}+\frac{1}{2}(\nu-\frac{B}{4})^{2}-\frac{\nu}{2}(\nu-\frac{B}{4})^{2}-\frac{\nu}{2}+\frac{\nu}{3}(\nu-\frac{B}{4})$} \\
\hline

\end{tabular}
\vspace{0.5 cm}
\caption{Some large $N$ limit evaluations of Wilson loops.}
\label{tab.gen}
\end{center}
\end{table}

Expressions for a generic model can be found in \cite{Minors}, reproduced here for the present model. Two observations then follow: 

\begin{enumerate}
\item For Wilson loops, they only depend on the positive Fourier coefficients of the potential. Hence, the model with an insertion $(\det(1+U))^{\nu}$ will give the same result as in our model, where the insertion is $(\det(1+U)\det(1+U^{\dag}))^{\nu}$.

\item For the more general case of two traces, they depend on both sets of coefficients, but still the result in the Table is simplified (with regards to the general one in \cite{Minors}) due to the symmetry $V(z)=V(z^{-1})$ of our model.

\end{enumerate}

\subsection{Large $N$ limits}

In section 3 we will be studying the large $N$ limit with fixed 't Hooft couplings,
where $A$ and $B$ scale to infinity like $N$. That is, a triple scaling limit. 
However, it is interesting to also study large $N$ limits with different scalings. We will now consider two cases: a) no scaling of the couplings and b) the double-scaling limit of the GWW model, where $\nu $ will be finite.

With this aim, we take  into account the equivalent Toeplitz determinant description given by \eqref{Toepl}. The strong Szegö theorem establishes that, in the large $N$ limit, the determinant is, as we have already seen, governed by the Fourier coefficients of the potential.

\bigskip
\noindent{\it The large $N$ limit without scaling  of couplings ($\nu, B$ fixed)}
\medskip

The $\nu $-dependent term adds a zero to the GWW weight function, and hence we have to consider the Fisher-Hartwig extension of Szegö asymptotics, which for the GWW model gives
\begin{equation}
\lim_{N\rightarrow \infty }Z_{N}(\nu =0)=e^{\frac{B^{2}}{16}-\frac{BN}{2}}.
\end{equation}

This is the large $N$ behavior of the GWW without scaling. In the usual double-scaling behavior of the model, this corresponds to the weak-coupling phase.

The asymptotics of the full model is that of the GWW but with a single FH singularity. Note that the FH singularity is just a zero of the weight function $\omega \left( \varphi \right)$ of the model for $\varphi=\pi$. Using the extended formula \cite{DAK2}, we obtain:


\begin{equation}
\lim_{N\rightarrow \infty }Z_{N}(\nu ,B)=\frac{e^{\frac{B^{2}}{16}-\frac{BN}{2}-\frac{\nu B}{2}} N^{\nu^{2}}G^{2}(1+\nu)}{2^{2N\nu }G(1+2\nu )}.
\end{equation}


The term that mixes the two parts of weight function is obviously (and simply) $\exp(-\frac{\nu B}{2})$. The term $N^{\nu ^{2}}G^{2}(1+\nu )/G(1+2\nu )$ is the large $N$ limit of the finite $N$ result for the pure FH singularity \eqref{eq:exactZEt1}.

\bigskip
\noindent{\it Large $N$ limit with fixed $\nu $ and fixed $B/N$}
\medskip

It is convenient to introduce a coupling $g^2$ as $B=-4/g^2$, to match Gross-Witten notation. We will also omit the multiplicative constant $e^{-NB/2}$ in \eqref{moregeneral}.
We would now like to investigate the theory in the large $N$ limit with fixed $\nu =A-N$ and fixed $\lambda \equiv g^2 N$.


We can study this limit by computing the average of the term $\left[ \cos (\theta/2) \right] ^{2\nu }$ in the two phases of the GWW matrix
model, described by the density of states of the model \cite{Gross:1980he}, and also given below. Expressing the power of the cosine in terms of a sum of multiple-angle cosines:
\be
\cos^{2\nu } \Big( \frac{\theta}{2} \Big) =\frac{1}{2^{2\nu }}\binom{2\nu}{\nu }+\frac{2%
}{2^{2\nu }}\sum_{k=0}^{\nu -1}\binom{2\nu }{k}\cos \left( \left( \nu -k\right)
\theta \right) ,
\label{cosexp}
\ee
we can then use a non-trivial but well-known result available for averages with $\rho(\phi)$ in the GWW model, \cite{Gross:1980he,Rossi:1996hs,Okuyama:2017pil,Alfinito:2017hsh}:
\begin{eqnarray}
W_k &=& \int_{-\phi_c}^{\phi_c}\dd\phi \cos k\phi\,\rho(\phi) 
\nonumber \\ &=&
\left\{
\renewcommand\arraystretch{1.3}
\begin{array}{l@{\quad}l}
0 \,, & \lambda\ge2,  \\
\displaystyle \left(1 - {\lambda\over2}\right)^{\!\!2}
{1\over k-1}\,P^{(1,2)}_{k-2}\!\left(1 - {\lambda}\right), &
\lambda\le2 \,, 
\end{array}
\right.
\end{eqnarray}
for $k \ge 2$ and where $P^{(\alpha,\beta)}_k$ are the Jacobi polynomials. These quantities $W_k$ are the multiply winded (with $k$-winding) GWW Wilson loop averages, that is:
\be
W_{k}=\frac{1}{N}\left\langle \mathrm{Tr}\ U^{k}\right\rangle. 
\ee
These Wilson loops have been studied in much more detail recently \cite{Okuyama:2017pil,Alfinito:2017hsh}. In particular, series expansions in $1/N$ have been proposed for both the purely perturbative genus expansion as well as for the instanton parts of the $W_k$. The case $w_{1}$ is of course the standard Wilson loop in the fundamental representation, which can also be obtained  
from the derivative of the GWW free energy:
\begin{equation}
W_1 =\frac{\lambda^2}{2N^2}\frac{\partial F}{\partial \lambda}\ , \quad F\equiv -\ln Z=
\left\{
\renewcommand\arraystretch{1.3}
\begin{array}{l@{\quad}l}
-\frac{N^2}{\lambda^2} \,, & \lambda\ge2 \,, \\
\displaystyle -N^2\left(\frac{2}{\lambda}+\frac12 \ln \frac{\lambda}{2}-\frac34\right) \,, & \lambda\le2 \,.
\end{array}
\right.
\label{w1-UN}
\end{equation}

In this way, we compute the partition function of our model in this limit in terms of a sum of winding Wilson loops of the GWW theory, in the two phases. From \eqref{cosexp}, we obtain
\be
\frac{Z(\lambda ,\nu )}{Z(\lambda,\nu=0)}=\langle  \cos^{2\nu } \Big(  \frac{\theta}{2}  \Big) \rangle =\frac{1}{2^{2\nu }}\binom{2\nu }{\nu }+\frac{2}{2^{2\nu }}%
\sum_{k=0}^{\nu -1}\binom{2\nu }{k}W_{\nu -k}(\lambda ) ,
\ee
where $Z(\lambda,\nu=0)=Z$ is the partition function of GWW model given in \eqref{w1-UN}.

In the strong-coupling (ungapped) phase we only have contribution of the fundamental Wilson loop and in the gapped phase, all winding wilson loops up to the $\nu$-nth winding contribute. All together, for this phase we have an expression for the partition function in terms of sums of Jacobi polynomials.
For the strong-coupling (ungapped) phase, $\lambda \geq 2$ this is:%
\be
\frac{Z(\lambda ,\nu )}{Z(\lambda,\nu=0)} =\frac{1}{2^{2\nu }}\left[ \binom{2\nu }{\nu }+\binom{2\nu }{%
\nu -1}\frac{2}{\lambda }\right],
\ee
and for the weak coupling phase, $\lambda \leq 2$ we have:
\be
\frac{Z(\lambda ,\nu )}{Z(\lambda,\nu=0)} =\frac{1}{2^{2\nu }}\left[ \binom{2\nu }{\nu }+2\left( 1-%
\frac{\lambda }{4}\right) \binom{2\nu }{\nu -1}+2\left( 1-\frac{\lambda }{2}\right)^{2} \sum_{k=0}^{\nu -2}\binom{2\nu }{k}\frac{P_{\nu -k-2}^{(1,2)}(1-%
\frac{\lambda }{2})}{\nu -k-1}\right].
\ee
It would be interesting to study how other results in \cite{Alfinito:2017hsh} for the $W_k$ can be generalized to our model.

\bigskip

\section{Large $N$ partition function in deformed GWW model}

Our starting point is the partition function \eqref{moregeneral}.
It can also be written as 
\begin{equation}
Z_{N}= \frac{1}{N!}\int_{(0,2\pi ]^{N}}\prod_{1\leq j<k\leq N}\left\vert e^{\ii\varphi
_{j}}-e^{\ii\varphi _{k}}\right\vert ^{2}\ \  \prod_{j=1}^{N}\exp \left( \frac{2}{g^2}\cos \left( \varphi
_{j}\right) +2\nu \ln \cos\frac{\varphi _{j}}{2} \right) \frac{\dd\varphi _{j}}{2\pi} \ .
\label{moregener}
\end{equation}%
As in the previous section, we have defined $B=-4/g^2$, to match Gross-Witten notation.
We have also dropped the overall constant factor $e^{-\frac{NB}{2}}$, which is not essential
for the analysis of the dynamics of the model (one may define the model without this factor).
As pointed out above, for $\nu=0$ the theory reduces to the GWW model.

In this section we study the large $N$ limit of the model with fixed 't Hooft couplings.
We consider the new variables
\begin{equation}
\lambda \equiv g^2 N\ ,\qquad \tau \equiv \frac{\nu }{N}=\frac{A}{N}-1\ ,
\end{equation}%
and take the limit $g\to 0$, $\nu \to \infty$, $N\to \infty$
with $\lambda $, $\tau $ fixed. This gives the planar limit of the theory.

In the GWW model, the physical range for the coupling $\lambda $ is $\lambda >0$. The range $\lambda<0 $ is formally equivalent to the range $\lambda>0 $ as one can switch the sign of $\lambda $ by a shift of integration variables $\varphi_j\to \varphi_j+\pi $. 
The equivalence no longer holds
in the present theory with the $\tau $ deformation, since such shift turns $\ln |\cos(\varphi/2)|$ into $\ln |\sin(\varphi/2)|$.  By virtue of this property, the present model with $\lambda<0$ can be alternatively viewed
as the usual GWW model with $\lambda>0$, but deformed by a term
$2\nu \ln |\sin\frac{\alpha} {2}|  $ instead of $2\nu \ln |\cos\frac{\alpha} {2} | $.
Therefore the region $-\infty <\lambda<0 $ has a new dynamics, which has to be investigated.

Regarding the range of the $\tau $ parameter, we note that in the finite $N$ partition function
convergence requires $\nu >-1/2$. In turn, this means positive $\tau $, as only $\nu \to +\infty$ would be allowed
on the basis of convergence. Nevertheless, as well known in matrix models, in the large $N$ double scaling
limit one can as well study potentials which are unbounded from below.
Thus we shall also  study the theory in the region of negative $\tau $, even though
this region may be  irrelevant for the interpretation of the model as a deformation of lattice gauge theory 
(the logarithmic singularity of the potential appearing at negative $\tau $ is a familiar one, extensively studied in Penner's matrix models).

\medskip
The partition function can be computed as usual by the method of saddle point.
Introducing a density of eigenvalues $\rho(\alpha)$ normalized to 1,
the saddle-point equations lead to the following integral equation
\begin{equation}
\frac{2}{\lambda}\,\sin\alpha + \tau\tan\frac{\alpha}{2} =P\int_{L}
d\beta\, \rho(\beta )\cot\left( \frac{\alpha-\beta}{2}\right)\ ,
\end{equation}
where $L$ represents the region where eigenvalues condense.
There are different solutions, thus different phases, according to the values
of the couplings $\lambda,\, \tau$.
The particular case $\tau=0$ corresponds to the GWW model.
In this case, for $\lambda<2$ the solution is given by
\begin{equation}
    \rho(\alpha ) = \frac{2}{\pi\lambda} \cos\frac{\alpha}{2} 
 \sqrt{m-\sin^2\frac{\alpha}{2}} \ , \qquad \alpha\in (-\alpha_0,\alpha_0)\ ,
\end{equation}
where $m\equiv\sin^2\alpha_0/2=\lambda/2$. This represents the ``gapped" phase. The cut exists for $|\alpha_0|<\pi$, {\it i.e.}  $m\leq 1$, which implies that, when $\tau=0$, this phase only exists in the weak coupling regime $0<\lambda\leq 2$. On the half-line $\{\tau=0, \lambda>2\}$, one has the ungapped (strong coupling) phase where the solution is given by
\be
\rho_{\rm ungap}=\frac{1}{2\pi}+\frac{1}{\pi \lambda}\, \cos\alpha \ ,\qquad  -\pi<\alpha\leq \pi\ .
\label{ungapGW}
\ee
\medskip

When $\tau\neq 0$, one can find the exact  one-cut solution
representing a gapped phase with $\alpha\in (-\alpha_0,\alpha_0)$
by standard methods, where the new $\alpha_0$ depends on $\tau$ and $\lambda $. The general solution is given by
\begin{equation}
    \rho(\alpha ) =\left( \frac{2}{\pi\lambda} \cos\frac{\alpha}{2} +\frac{\tau}{2\pi}\frac{1}{\cos\frac{\alpha_0}{2}\cos\frac{\alpha}{2}}
\right) \sqrt{m-\sin^2\frac{\alpha}{2}}\ .
\label{soliro}
\end{equation}
In terms of the variable $t=\sin\frac{\alpha}{2}$, the eigenvalue density becomes
\begin{equation}
    \rho(t) =\left( \frac{4}{\pi\lambda} +\frac{\tau}{\pi}\frac{1}{\sqrt{1-t_0^2}\, (1-t^2)}
\right) \sqrt{t_0^2-t^2}\ ,\qquad t_0^2=m=\sin^2\frac{\alpha_0}{2}\ .
\end{equation}
where the transformation of the measure was taken into account.
The width of the eigenvalue distribution is determined by the normalization condition.
Integrating \eqref{soliro}, we find
\begin{equation}
1=\int_{-\alpha_0}^{\alpha_0}
d\beta\, \rho(\beta )=\frac{2m}{\lambda}
+\left(\frac{1}{\sqrt{1-m}}-1\right) \tau\ .
\label{normaa}
\end{equation}
This leads to a cubic equation for $m$ which determines $m=m(\lambda,\tau)$:
\be
m^3 - m^2 (1+\lambda  (\tau +1))+\frac{1}{4} m \lambda   (\tau +1) (\lambda  \tau +\lambda +4)-\frac{1}{4} \lambda ^2 (2 \tau +1)=0\ .
\ee
The phase can exist as long as $0<m\leq 1$ and as long as $\rho$ is non-negative in the interval $(-\alpha_0,\alpha_0)$.
The conditions for the existence of the phase can be more simply elucidated  by solving (\ref{normaa}) for $\tau $.
We find
\begin{equation}
\tau = \frac{\sqrt{1-m}\ (\lambda -2 m)}{\lambda  \left(1-\sqrt{1-m}\right)}\ .
\label{taw}
\end{equation}

\subsection{Case $\lambda >0 $}

The behavior of $\tau $ is shown in figs 1a, 1b, corresponding to the cases
$\lambda\geq2$ and $0<\lambda<2$.
In either case, as $m\to 0^+$, one has $\tau\approx \frac{2}{m}\to \infty$.

\begin{figure}[h!]
\centering
\begin{tabular}{cc}
\includegraphics[width=0.4\textwidth]{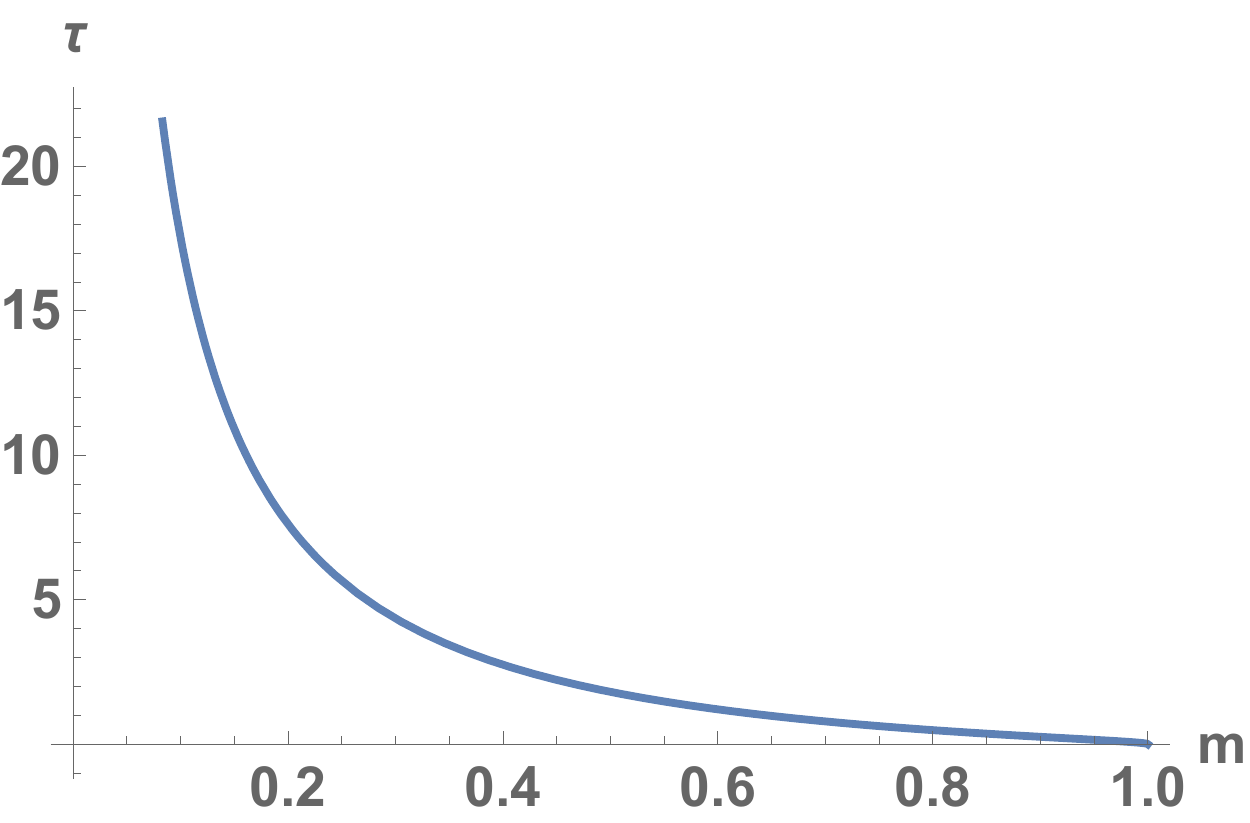}
&
\qquad \includegraphics[width=0.4\textwidth]{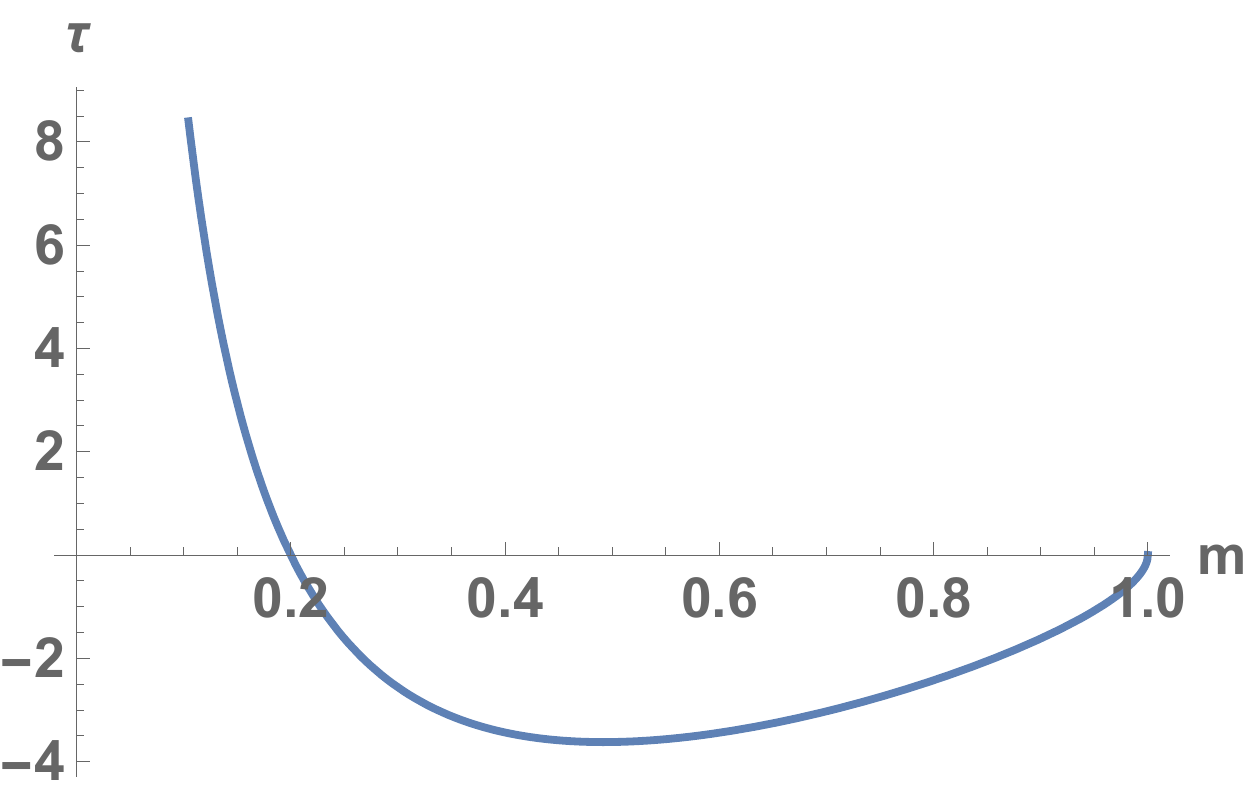}
\\
(a)&(b)
\end{tabular}
\caption
 {The relation between the width of the eigenvalue distribution and the coupling $\tau$ for given $\lambda $. In the interval $0<\lambda<2$,  $\tau (m) $ 
 has the form of figure (b). (a) $\lambda=4$. (b) $\lambda=1$.}
\label{tauab}
\end{figure}

Consider first $\lambda\geq 2$. Then
$\tau $ is monotonic decreasing until $\tau=0$ at $m=1$.
For any given $\lambda\geq 2$ and $\tau$, there is a unique $m$,
hence a unique normalizable solution.

On the other hand, if $0<\lambda<2$, $\tau $ decreases up to a minimum value
$\tau_{\rm min}<0$ and then increases monotonically until
$\tau=0$ at $m=1$.
In the region where $\tau >0$,  there is a unique solution for $m$ for any given 
$0<\lambda<2$. 
On the other hand, in the region $\tau_{\rm min}<\tau<0$, there are
two solutions for $m$ for a given $\tau$. However, the solution in the branch with
higher $m$ is ruled out because $\rho $ becomes negative in two intervals
$(\alpha_1,\alpha_0)$ and $(-\alpha_0,-\alpha_1)$ near the endpoint of the 
eigenvalue distribution.

One surprising feature is that, when $\tau>0$, the gapped phase 
represented by the solution \eqref{soliro} exists
all the way from $\lambda=0$ to $\lambda=\infty$. The reason can be understood from the potential, which goes to infinity as  $\alpha =\pm \pi $ is approached
(see fig. \ref{potRI}).
This feature keeps the eigenvalues confined in a finite domain with $|\alpha_0|<\pi$ for
all couplings $0<\lambda <\infty$ as long as $\tau>0 $.
As a result, the GWW phase transition disappears from the region $0<\lambda <\infty$
when a positive coupling $\tau $ is turned on. In other regions of the $(\lambda,\tau)$ parameter space there will be other phase transitions of different nature.

\begin{figure}[h!]
\centering
\includegraphics[width=0.55\textwidth]{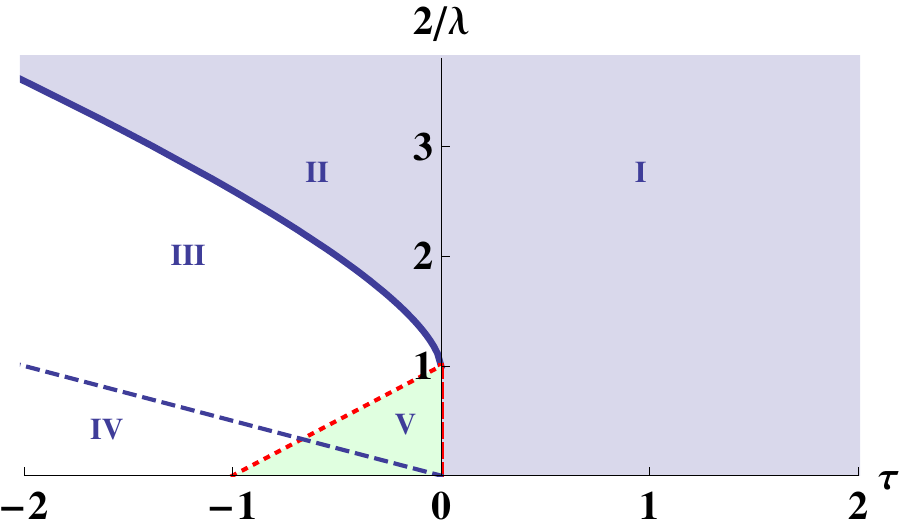}
\caption{Partial representation of the phases of the theory in the region $\lambda >0$. The one-cut solution \eqref{soliro} exists in the shaded regions I and II. In region IV the minimum of the potential at $\alpha=0$ has disappeared. The ungapped solution \eqref{singap} exists in region V. It matches continuously on the red dashed vertical line with the solution at  $\tau\geq 0$. 
}
 \label{phasediagram}
\end{figure}

\begin{figure}[h!]
\centering
\includegraphics[width=0.45\textwidth]{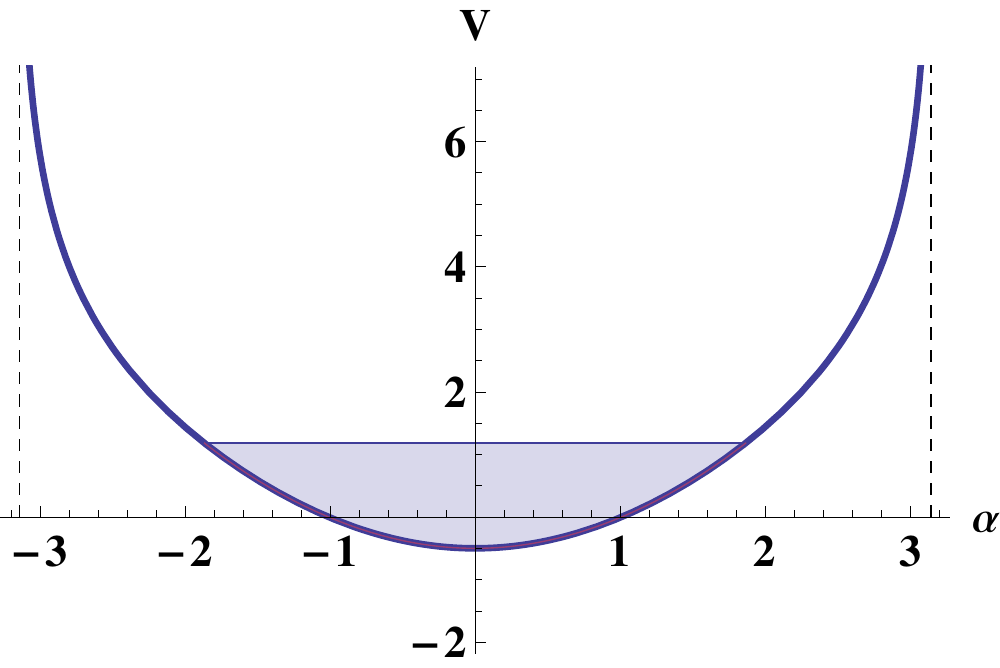}
\caption{Potential in region I (here $\lambda=4$, $\tau=1$).}
 \label{potRI}
\end{figure}

The transition to the region $\tau<0$ depends on the value of $\lambda$. For $0<\lambda<2$, one has $m<1$ so $\alpha_0<\pi$ (see fig. \ref{tauab}b); 
the solution \eqref{soliro} has no discontinuity in going from $\tau>0$ to $\tau<0$, {\it i.e.} from region I to region II of fig. \ref{phasediagram}.
On the other hand, for $\lambda>2$,
in the region $\tau\to 0^+$, one has $m\to 1^-$ and
$\tau \approx \frac{\lambda-2}{\lambda}\, \sqrt{1-m}=\frac{\lambda-2}{\lambda}\, \cos\frac{\alpha_0}{2}$.
The solution \eqref{soliro} 
becomes
\begin{equation}
    \rho\to \frac{1}{2\pi}+\frac{1}{\pi \lambda}\, \cos\alpha  \  ,\qquad \lambda>2,\ \tau\to 0^+\ .
\end{equation}
Thus it matches continuously 
the ungapped solution \eqref{ungapGW} of the GWW model that exists on the half-line $\{\tau=0, \lambda>2\}$.

When $\tau $ is negative, the solution \eqref{soliro}  exists
in the shaded region II of fig. \ref{phasediagram}, $\tau_{\rm min}(\lambda)<\tau<0$ and $0<\lambda<2$, where
$\tau_{\rm min}(\lambda)$ is the minimum of $\tau$ for a given $0<\lambda<2$ (fig. \ref{tauab}b). 
The curve separating regions II and III is $\tau_{\rm min}(\lambda)$ and it can be represented parametrically as follows 
\begin{eqnarray}
&&\tau= \frac{2 (1-m)^{\frac{3}{2}}  }{2-3m -2 (1-m)^{\frac{3}{2}}}  \nonumber\\
\\
&& \lambda=6m +4(1-m)^{\frac{3}{2}} -4\ ,\qquad 0 <m<1\ .
\end{eqnarray}
On the critical curve, one has $m \to 1$ for $(\tau,2/\lambda)\to (0^-,1)$. In the opposite limit, one has $m\to 0$
for $ (\tau,2/\lambda)\to (\infty,\infty)$.

Because $m$ never reaches 1 for $\tau< 0$ on the critical curve ({\it i.e.} $\alpha_0<\pi$), there cannot be a continuous transition to an ungapped phase
across the critical line separating regions II and III.
There is an ungapped phase in the GWW line at $\tau=0$ and $\lambda>2$,
and also in region V. The solution is constructed below.

\begin{figure}[h!]
\centering
\begin{tabular}{cc}
\includegraphics[width=0.4\textwidth]{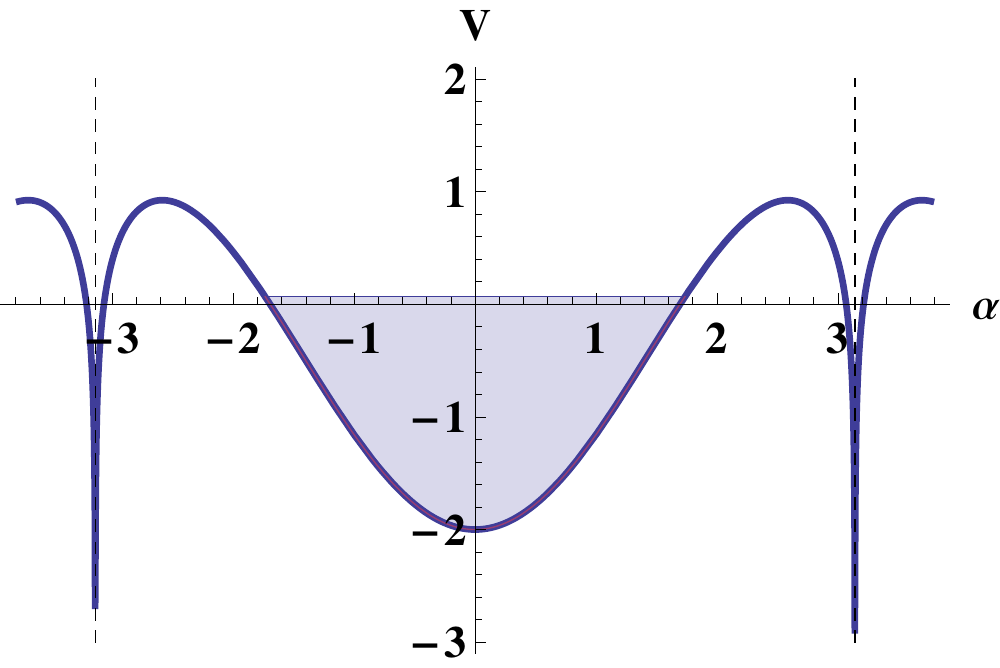}
&
\qquad \includegraphics[width=0.4\textwidth]{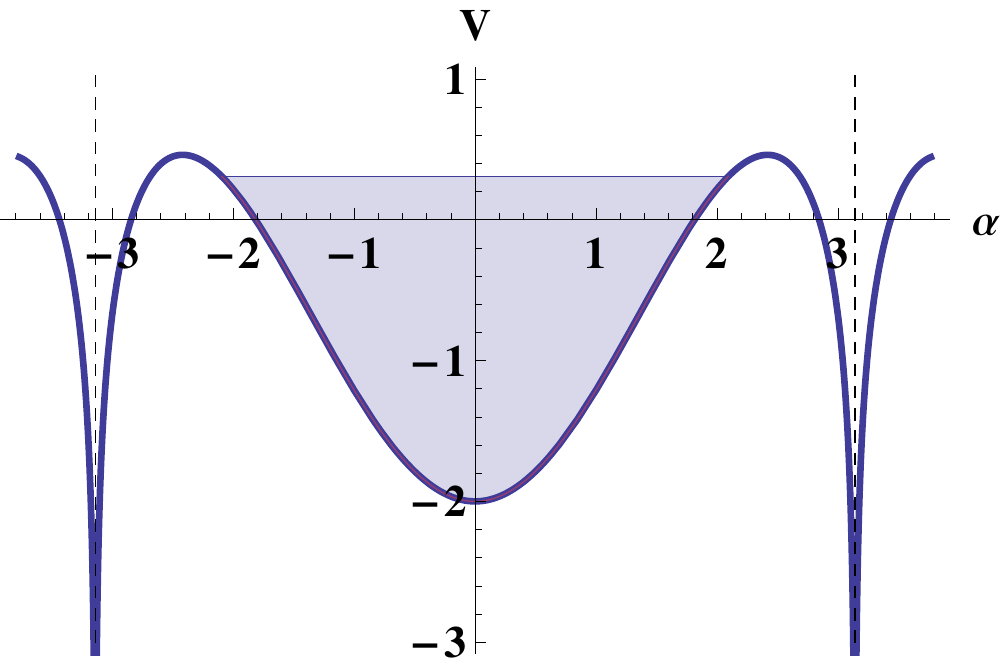}
\\
(a)&(b)
\end{tabular}
\caption
 {Potential.  
 (a) $\tau=-0.3, \lambda=1$, corresponding to region  II. (b) $\tau=-0.5, \lambda=1$, $\alpha_0=2\pi/3$, corresponding to the critical line separating region II and region III.}
\label{pot2c}
\end{figure}

\begin{figure}[h!]
\centering
\begin{tabular}{cc}
\includegraphics[width=0.4\textwidth]{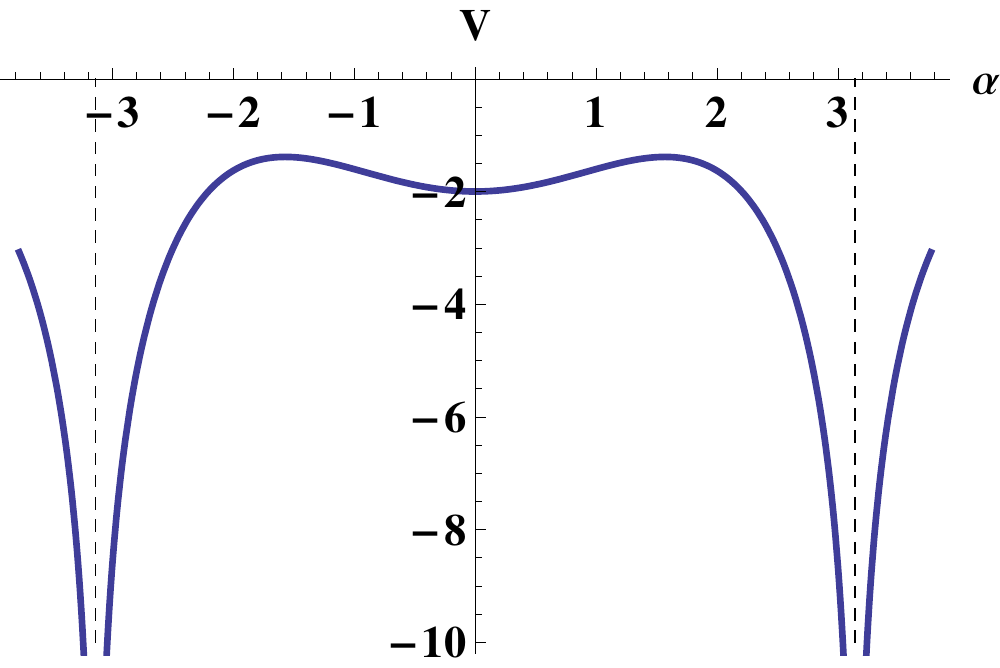}
&
\qquad \includegraphics[width=0.4\textwidth]{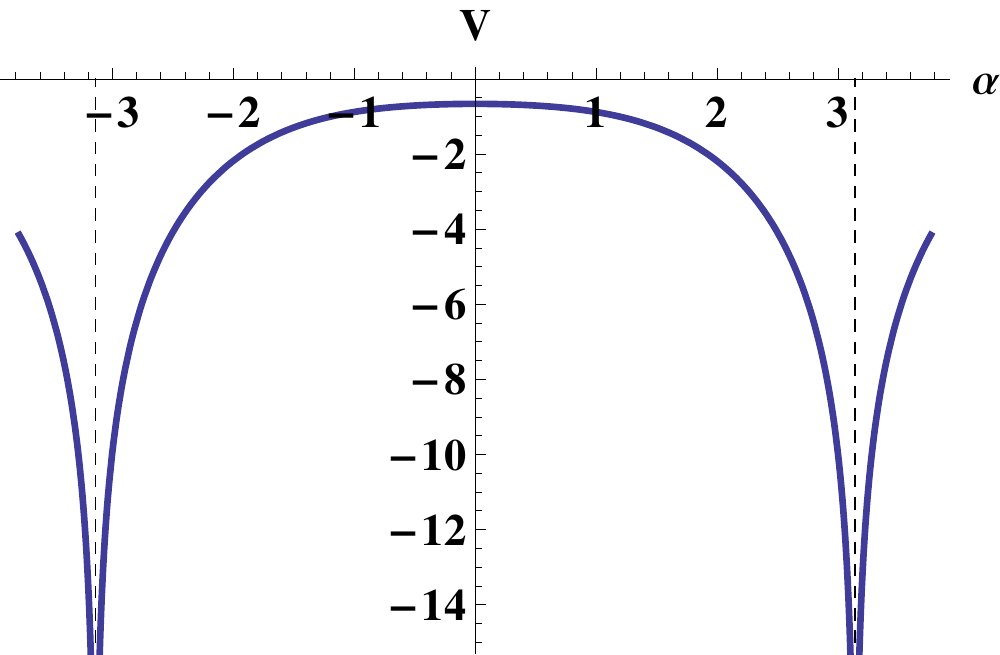}
\\
(a)&(b)
\end{tabular}
\caption
 {Potential.  
 (a) $\tau=-2$, $\lambda=1$ corresponding to region III. (b) $\tau=-2$, $\lambda=3$, corresponding to region IV.}
\label{pot34}
\end{figure}

The phase diagram of fig. \ref{phasediagram} summarizes the 
different regimes of the theory with $\lambda>0$.
An interesting question is what is the solution in regions III and IV. The new solution must match continuously the
solution \eqref{soliro}  on the critical line separating regions II and III. 
A hint on the nature of the new phases is provided by the form of the potential in the different regions:

\begin{itemize}
    
    \item In region I, the potential is shown in fig. \ref{potRI}.
    It goes to infinity as  $\alpha=\pm\pi$ is approached, which explains
    why for $\tau>0$ there is never a transition to an ungapped solution.

\item In region II, the potential has the form of fig. \ref{pot2c}a.
As the critical line is approached, eigenvalues almost fill the well, indicating a possible transition to a solution
where part of the eigenvalues sit on the wells at $\pm\pi$. 

\item Finally, figs. \ref{pot34}a, b represent the potential in regions III and IV. They suggest the existence of a critical line
near the dashed line in fig. \ref{phasediagram} where there is a transition  to a solution
where {\it all} eigenvalues sit on the wells at $\pm\pi$.

\end{itemize}

Let us now describe the 
 ungapped solution where eigenvalues are spread  over the whole circle, $\alpha_0=\pi$.
 The normalized solution is
given by
\begin{equation}
    \rho(\alpha ) =\frac{(1+\tau)}{2\pi} +\frac{1}{\pi\lambda }\cos\alpha  
    -\frac12 \tau \left(\delta(\alpha-\pi )+
     \delta(\alpha+ \pi )\right)\ .
\label{singap}
\end{equation}
The solution continuously matches the GWW solution \eqref{ungapGW} on the half-line $\tau=0$, $\lambda>2$,
which also matches the solution \eqref{soliro} in region I at $\tau\to 0^+$, $\lambda>2$.
The critical line separating the region where this solution exists can be found from the requirement that $\rho$ is non-negative
in the whole interval $[-\pi,\pi)$. This gives the condition
\be
\tau+1 > \frac{2}{\lambda} > -\tau -1\ .
\ee
The complete relevant region  for this phase is shown in fig. \ref{phaselam}.

\medskip

\subsection{Case $\lambda <0 $} 

\begin{figure}[h!]
\centering
\includegraphics[width=0.45\textwidth]{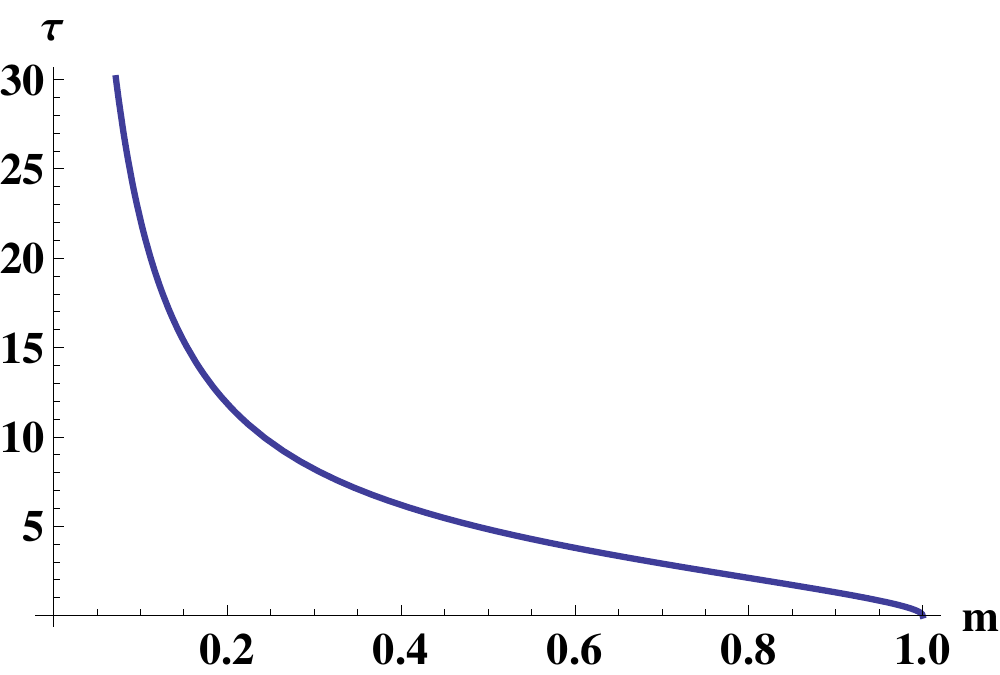}
\caption{$\tau $ in terms of $m$ given by \eqref{taw} (here $\lambda=-1$). For any given $\tau>0$ and $\lambda<0$ there is a unique solution $m$ to the normalization condition.}
 \label{normanegativo}
\end{figure}

The solution  \eqref{soliro} still holds in part of the region $\lambda <0 $, $\tau>0$.
For all $\lambda <0$, provided  $\tau>0$, there is a unique solution $m(\lambda,\tau)$ to the normalization condition for each pair $(\lambda,\tau )$. This is shown in fig. \ref{normanegativo}.
On the other hand, there is no solution for $\tau<0$, so the phase described by
the solution \eqref{soliro} does not exist in the quadrant $(\lambda<0,\tau<0)$.
In this quadrant eigenvalues should condense on a cut near the wells at $\pm \pi $.
We expect a smooth transition from region IV of fig. \ref{phasediagram}, since the potential is qualitatively the same.

\begin{figure}[h!]
\centering
\begin{tabular}{cc}
\includegraphics[width=0.4\textwidth]{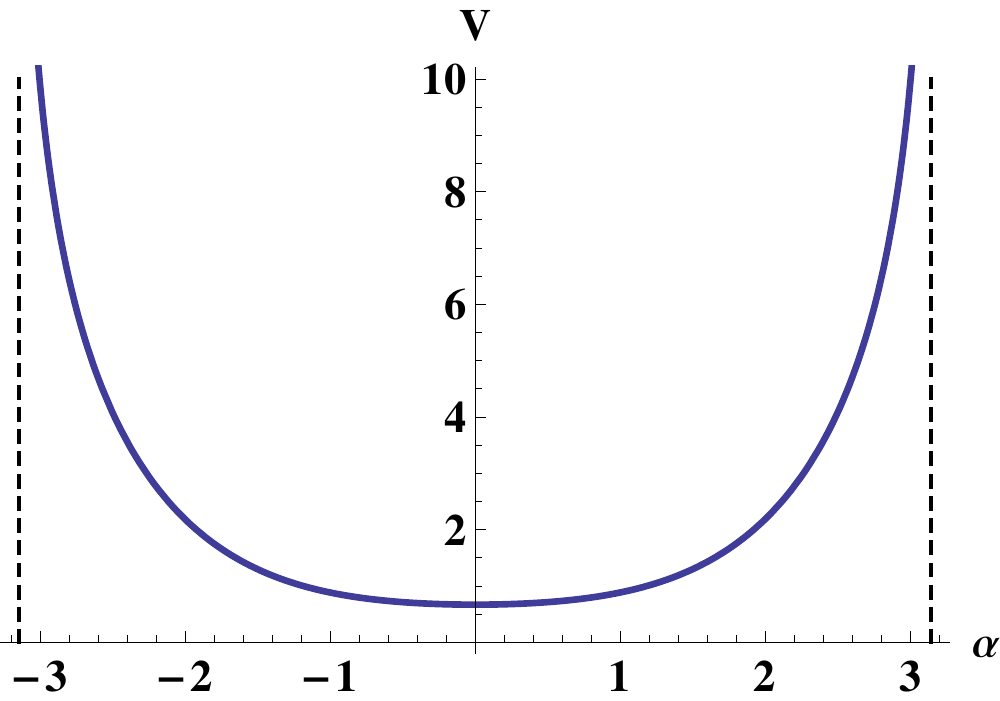}
&
\qquad \includegraphics[width=0.4\textwidth]{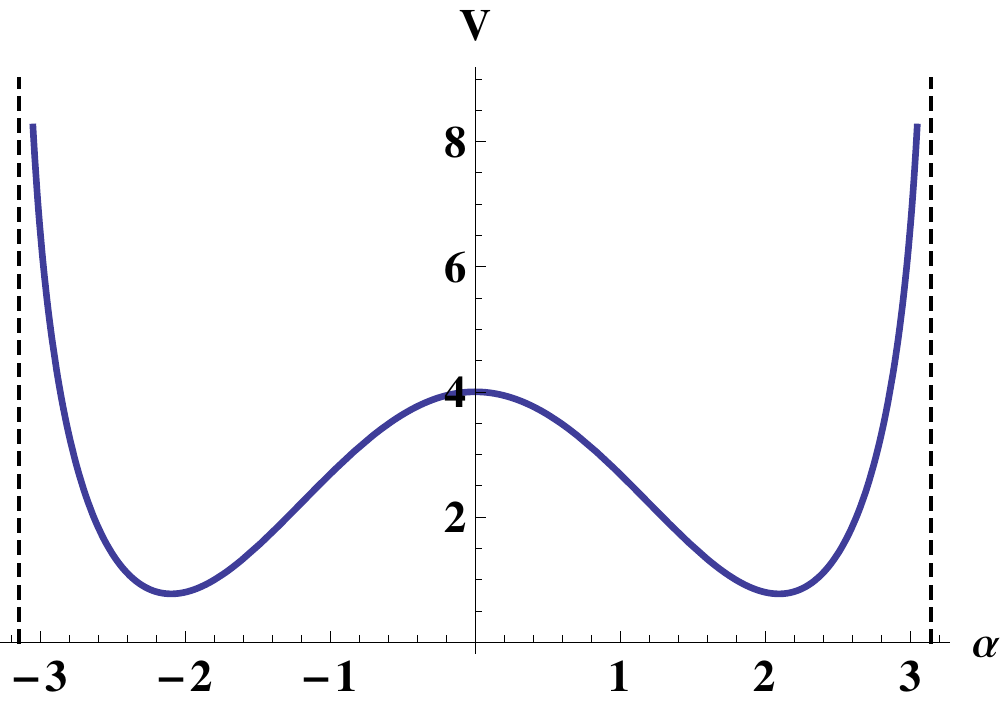}
\\
(a)&(b)
\end{tabular}
\caption
 {Potential.  
 (a) $\tau=2, \lambda=-3$, corresponding to region VI with $\lambda<\lambda_{\rm cr}$. (b) $\tau=2, \lambda=-0.5$,  corresponding to  region VII with $\lambda>\lambda_{\rm cr}$.}
\label{potenciales}
\end{figure}

Thus, let us consider the quadrant $(\lambda<0,\tau>0)$.
Starting with large $|\lambda |$, the potential has an absolute minimum at 
$\alpha = 0$ and it develops a double well for a  coupling $\lambda>\lambda_1$,  $\lambda_1=-4/\tau$ (see fig. \ref{potenciales}).
Therefore one expects a phase transition where the eigenvalue distribution splits in two
cuts, at some critical coupling $\lambda_{\rm cr} >\lambda_1$ ({\it i.e.} $|\lambda_{\rm cr}|< |\lambda_1| $).
This phase transition is nothing but the unitary model version of the phase transition
described in \cite{Russo:2020pnv} for the corresponding Hermitian matrix model \eqref{partis}.\footnote{Note the difference in notation. The parameters 
$(\lambda,\tau)$ corresponds to
$(-4/\kappa,\lambda)$ in  \cite{Russo:2020pnv}. The study of the Hermitian
model carried out in  \cite{Russo:2020pnv} only covers the region  
$\tau\geq 1$ (which in the unitary model corresponds to the region $\tau\geq 0$ due to the contribution to the Jacobian from the stereographic map, that shifts $A$ to $A-N$).}
The eigenvalue density may be found from the two-cut solution in  \cite{Russo:2020pnv} by using the duality map between Hermitian and unitary models
\cite{Mizoguchi:2004ne} (see also \cite{Okuyama:2017pil}). However, in this paper we will not explore this region.

\begin{figure}[h!]
\centering
\begin{tabular}{cc}
\includegraphics[width=0.4\textwidth]{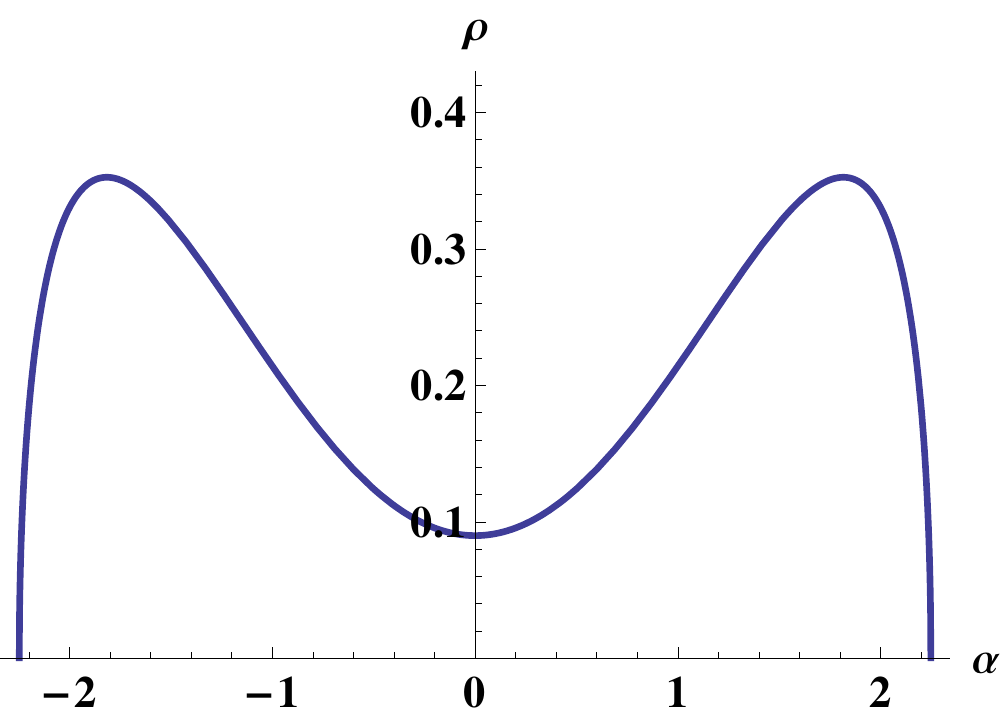}
&
\qquad \includegraphics[width=0.4\textwidth]{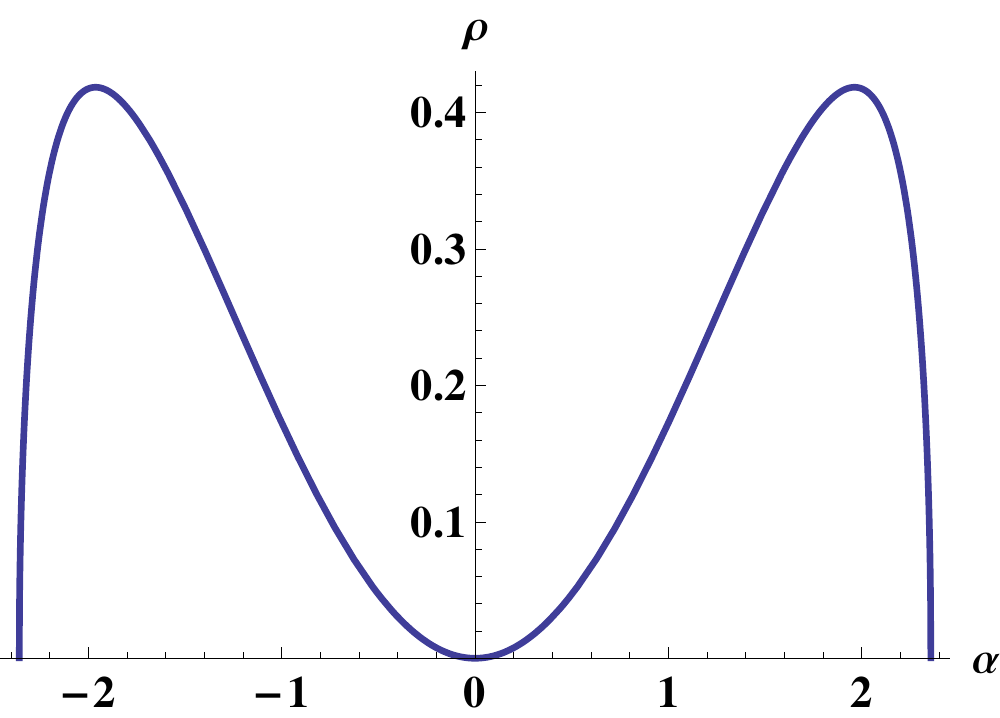}
\\
(a)&(b)
\end{tabular}
\caption
 {Eigenvalue density.  
 (a) $\tau=2, \lambda=-1$, corresponding to region  VI. 
 (b) $\tau=2, \lambda=\lambda_{\rm cr}(\tau=2)$,  corresponding to the critical line. For higher $\lambda$, the eigenvalues split in a two-cut distribution.}
\label{densidades}
\end{figure}

Like in \cite{Russo:2020pnv}, the critical line occurs when the eigenvalue density vanishes at $\alpha=0$ (see fig. \ref{densidades}b).
This gives the equation
$$
\lambda \tau =-4\sqrt{1-m}\ .
$$
Hence $|\lambda_{\rm cr}/\lambda_1|<1$, which means that the phase transition does not occur immediately when the double well forms, but for a greater $\lambda $. When $\lambda$ overcomes $\lambda_1=-4/\tau$, there is still a one-cut solution
due to overfilling of eigenvalues fully  covering the two wells. At $\lambda>\lambda_{\rm cr}$,
the eigenvalues get separated in two sets filling part of each well. 
Combining with \eqref{taw} --~which implicitly determines $m$ in terms of $\lambda, \ \tau$, we find
the critical line
\be
\lambda_{\rm cr} = -\frac{4}{\tau ^2}\left(\tau -\sqrt{2 \tau +1}+1\right)\ .
\label{lamcri}
\ee
For  $\tau \to \infty $, one has $m\to 0$ and $\lambda \approx -4/\tau $.
In the opposite limit, for $\tau \to 0^+$, one has $m\approx 1$ and $\lambda_{\rm cr}\to -2$. This is expected, since, as explained above,
when $\tau=0$ the theory with $\lambda<0$
is equivalent to the theory in the interval  $\lambda>0 $ and thus it has
a GWW phase transition at $\lambda =-2$.

Viewing the $\lambda<0$ model as the standard GWW model with $\lambda>0$ deformed by
$2\tau \ln |\sin\frac{\alpha}{2}|$, one can get another insight on the origin of the two-cut solution.
This deformation makes the potential to go to $+\infty$ as $\alpha\to 0$; as a result,
the eigenvalue distribution --~that for $\tau=0$ and small $\lambda $   would have support in a cut near $\alpha=0$~-- now necessarily splits into two cuts. On the other hand, for strong $\lambda $,
the eigenvalue distribution, that in the $\tau=0$ theory would cover the whole circle, now must have a gap near $\alpha =0$ due to the infinite wall.
Therefore, in the GWW model with $\lambda>0$ deformed by
$2\tau \ln |\sin\frac{\alpha}{2}|$ the GWW phase transition extends to the full critical line $|\lambda_{\rm cr}(\tau)|$.
For any $\tau>0$, the phase transition now involves a transition between a solution with two gaps and a solution with one gap.

\begin{figure}[h!]
\centering
\includegraphics[width=0.55\textwidth]{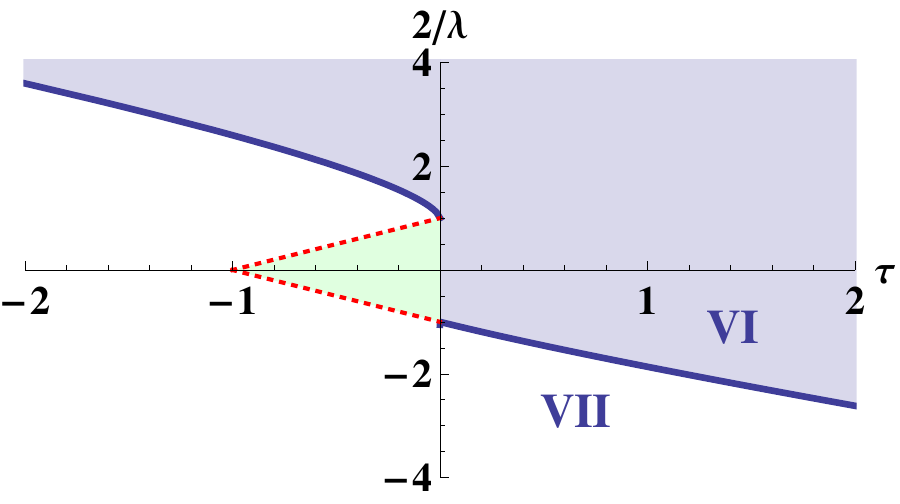}
\caption{Phase diagram.
The gapped phase described by solution \eqref{soliro} exists in the grey shaded region. The green shaded area describes the region V where the ungapped phase solution \eqref{singap}  exists. Below the critical line $\lambda_{\rm cr}$ in  region VII there is a two-cut phase where eigenvalues accumulate at the two minima of the potential ({\it c.f.} fig. \ref{potenciales}b). }
 \label{phaselam}
\end{figure}

The resulting phase diagram is shown in fig. \ref{phaselam}.


\subsection{Free energy and Wilson loops}

Here we use the eigenvalue densities \eqref{soliro} and \eqref{singap} to compute the free energy
in regions I, II, V and VI. By differentiating with respect to $\lambda $, one obtains a formula for the VEV of the Wilson loop:
\be
\langle W\rangle =\frac{\lambda^2}{2N^2}\frac{\partial F}{\partial \lambda}= \langle \cos\alpha\rangle \ .
\ee

Let us first consider the eigenvalue density \eqref{soliro}.
Computing the integrals we find 
\be
\langle W\rangle = \int_{-\alpha_0}^{\alpha_0} d\alpha\ \rho(\alpha)\  \cos\alpha\ =\frac{1}{\lambda}m (2-m)  +\tau \left(1 -\sqrt{1-m}\right)\ .
\ee
The  dependence of the Wilson loop as a function of the couplings
$\lambda $ and $\tau $ is complicated, because $m(\lambda,\tau)$ is a solution of a cubic equation. However, we can expand $\langle W\rangle$ in series to exhibit the small $\lambda $ and the small $\tau$ behavior.
At small $\lambda>0$, one has the expansion
\be
\langle W\rangle =1-\frac{\lambda }{4}+\frac{\lambda ^2
   \tau }{16}+\frac{1}{64} \lambda ^3
   \left(\tau -\tau
   ^2\right)+\frac{\lambda ^4 \left(4
   \tau ^3-12 \tau ^2+5 \tau
   \right)}{1024}+O\left(\lambda
   ^5\right)\ .
\ee
At small $\tau$ and in the region $0<\lambda<2$, one has the expansion
\be
m=\frac{\lambda }{2}+\frac{\lambda  \left(2-\lambda -\sqrt{4-2 \lambda }\right) \tau }{2 (2-\lambda)}+\frac{\left(2-\sqrt{4-2 \lambda }\right) \lambda ^2 \tau ^2}{4 (2-\lambda)^2}+O\left(\tau ^3\right)\ ,
\ee
and
\be
\langle W\rangle =1-\frac{\lambda
   }{4}+\left(2-\frac{\lambda
   }{2}-\sqrt{4-2 \lambda }\right)
   \tau +\frac{\lambda  \left(\lambda +2 \sqrt{4-2 \lambda }-4\right) \tau ^2}{4 (2-\lambda )} +O\left(\tau ^3\right)\ .
\label{sml}
\ee
The first term in \eqref{sml} is of course the VEV of the Wilson loop in the GWW model in the weak coupling phase. The second term displays the non-analytic behavior at the GWW critical  point $\lambda=2$.

Let us now consider the regions where $\lambda >2$. In this region the parameter $m$ solving the normalization condition \eqref{normaa} has the small $\tau $ expansion
\be
m=1-\frac{\lambda ^2 \tau ^2}{(\lambda -2)^2}+O\left(\tau   ^3\right)\ .
\label{mmm}
\ee
This gives  
\be
\langle W\rangle = \frac{1}{\lambda }+\tau -\frac{\lambda }{\lambda -2}\, \tau ^2+O\left(\tau   ^3\right)\ .
\label{delam}
\ee
For $\tau\to 0$, this agrees with the VEV of the Wilson loop of GWW theory in the strong coupling phase.

Another useful observable is
\be
-\frac{1}{N^2}\frac{\partial F}{\partial \tau}= \langle \ln \cos^2\frac{\alpha}{2}\rangle = \int_{-\alpha_0}^{\alpha_0} d\alpha\ \rho(\alpha)\   \ln \cos^2\frac{\alpha}{2}\, .
\ee
The order of a phase transition is determined by the continuity properties of the derivatives of the free energy in crossing a critical line
in $(\lambda,\tau)$ space. In particular, for a third-order phase transition, the second derivatives
$\partial^2_\tau F$, $\partial^2_\lambda F$ and $\partial_\lambda \partial_\tau F$
must be continuous. Hence this observable and $\langle W\rangle$ are instrumental in order to classify
the different types of transitions occurring in  $(\lambda,\tau)$ space.
Computing the integral, we obtain
\be
-\frac{1}{N^2}\frac{\partial F}{\partial \tau}=\frac{1}{\lambda}J_1 +\tau J_2\ ,
\ee
with
\begin{eqnarray}
&& J_1= 4-4 \sqrt{1-m}-2 m \left(1-2 \log
   \frac12 \left(1+\sqrt{1-m}\right)\right)\ ,
\nonumber\\
\\
&&J_2= -\left( \frac{1}{\sqrt{1-m}}+1  
  \right) \log \left(\frac{4 \left(2-m-2
   \sqrt{1-m}\right)}{m^2}\right)-\log
   (1-m)\ .
\end{eqnarray}
In the region $\lambda >2$, one finds the small $\tau $ expansion,

\begin{eqnarray}
-\frac{1}{N^2}\frac{\partial F}{\partial \tau} &=& \frac{2}{\lambda }-\log 4+\tau 
   \left(\log \frac{(\lambda
   -2)^2}{16\tau^2\lambda ^2}+2\right)
 \nonumber\\
  &+&\frac{\lambda  \tau ^2 (3\lambda 
   (1-\log 2)-6+10\log 2)}{(\lambda
   -2)^2}+O\left(\tau ^3\right)
\label{detau}
\end{eqnarray}
We can check that \eqref{delam} and \eqref{detau} satisfy the integrability condition $\partial_\lambda \partial_\tau F=\partial_\tau \partial_\lambda F$.

\begin{figure}[h!]
\centering
\includegraphics[width=0.55\textwidth]{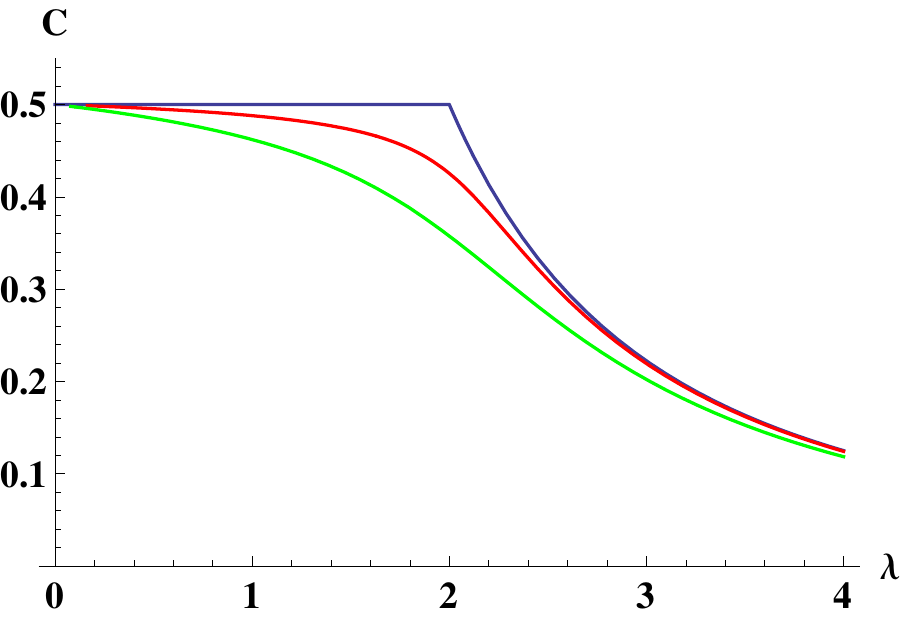}
\caption{Specific heat $C/N^2=-2\partial_\lambda W$ as a function of  $\lambda$. From top to bottom,  $\tau=0$, $\tau=0.03$ and $\tau=0.1$.}
 \label{calorespe}
\end{figure}

The critical behavior of the free energy can be studied by expanding these expressions  near critical lines.
As a first example, we consider the GWW phase transition. We have seen that
there is no critical line when $\tau>0 $ and that the phase transition becomes a cross-over. Figure \ref{calorespe} illustrates how the phase transition is smoothed out,
by means of a plot of the ``specific heat" $C=dE/d\lambda$, $E=\lambda^2 \partial F/\partial\lambda =2N^2\langle W\rangle$ in terms of the ``temperature" $\lambda$.
This shows that $\partial_\lambda^2 F$ is smooth for $\tau>0$.
A similar behavior is observed for $\partial_\tau\partial_\lambda F$ and $\partial_\tau^2 F$.

As a second example,  let us discuss
the phase transition across the critical line defined by $\{ \tau=0,\ \lambda>2\}$ in going from positive $\tau $ to negative $\tau$.
This is a rather drastic 
transition, since the potential has  infinite wells on the other side, in region V. 
Nevertheless, from a mathematical viewpoint, it is interesting to
 examine analytic properties in going across this line. We need the first derivatives of the free energy with respect to the couplings $\lambda$ and $\tau$ in region V,
where the eigenvalue density is given by \eqref{singap}.
In this phase the integrals are very simple and we obtain
\be
\langle W\rangle = \frac{1}{\lambda }+\tau\ ,
\ee
\be
-\frac{1}{N^2}\frac{\partial F}{\partial \tau}= \langle\ln \cos\frac{\alpha}{2}\rangle=\frac{2}{\lambda} -\log 4 -\tau \log 4
-\tau \ln \cos\frac{\pi-\epsilon}{2}\ .
\ee
We see that, in phase V,  the VEV of $\ln \cos\frac{\alpha}{2} $ has a logarithmic singularity, which
here has been regularized with a cutoff $\epsilon>0$.
Note also the crucial role of the $\delta$ function term in the density \eqref{singap}
to produce the $\tau $ term in $\langle W\rangle$, which is required by the integrability condition
$\partial_\lambda \partial_\tau F=\partial_\tau \partial_\lambda F$.
From the small $\tau $ expansions, we see that $\partial_\lambda F$ and $\partial_\tau F$
are continuous across the $\tau=0$ line. Likewise,   $\partial^2_\lambda F$ and $\partial_\lambda \partial_\tau F$ are also continuous.
However,  $\partial^2_\tau F$ is discontinuous. This indicates that the phase transition across the $\tau=0$ line in going from phase V to phase I (or to phase VI) is
second order.

One can similarly compute the VEV of winding Wilson loops $ W_k= \cos k\alpha$. 
In particular, in the gapped phase
described by \eqref{soliro}, we find
\begin{eqnarray}
&& \langle \cos 2\alpha\rangle = \frac{2 m (1-m)^2}{\lambda }+\left((1+m)\sqrt{1-m}
  -1\right) \tau\ ,
\nonumber\\
&& \langle \cos 3\alpha\rangle =  \frac{m (2-5 m) (1-m)^2}{\lambda }
-\left( (2m^2+1) \sqrt{1-m} -1\right) \tau\ ,
\end{eqnarray}
etc. Near the critical line $\{ \tau=0, \ \lambda>2\}$, one can use the expansion \eqref{mmm}
to find a general formula for the near critical behavior:
\begin{eqnarray}
 \langle \cos k\alpha\rangle =  (-1)^{k+1}\ \tau \left(1 - \frac{k \lambda }{(\lambda
   -2)} \tau +O\left(\tau ^2\right) \right)\ ,\qquad k=1,2,\cdots
\label{wku}
\end{eqnarray}
This can be compared with the  result in the ungapped phase, {\it i.e.} on the other side of the transition. Using \eqref{singap}, we find
\be
\langle W_k\rangle =  (-1)^{k+1}\ \tau \ .
\ee
Comparing with  \eqref{wku}, we see that all $ W_k$ and $\partial_\tau W_k$ are continuous as functions of $\tau $ at $\tau=0$, but the second derivatives
$\partial^2_\tau W_k$ are discontinuous.
Once again, note the crucial role of the delta function terms in \eqref{singap}, this time to ensure continuity of  $\partial_\tau W_k$.

\medskip

In conclusion, we have seen that the theory has a complicated phase structure.
It is clear that much deeper analysis is required to establish the dominant phases in each region and, in particular, to determine the order of phase transitions that occur when crossing the different critical lines.
If we restrict ourselves to the  ``physical" $\tau>0$ region, the phase structure seems to be simpler, with a single phase transition from region VI to region VII, which can be thought of as the extension of the GWW phase transition in the presence of the $\tau$ coupling (analogous to the liquid-vapor phase transition, in a phase diagram that includes temperature and pressure). The transition should be of the third order, being the counterpart of the third-order phase transition found in  \cite{Russo:2020pnv} in the dual Hermitian model.
We leave these problems for future work.

\section{Discussion}

The generalized GWW matrix theory studied in this work may be
viewed as two-dimensional lattice gauge theory with an insertion
\be
\hat O^\nu\ ,\qquad \hat O\equiv \frac{1}{4}\det(1+U)\det(1+U^\dagger)\ , 
\label{insertion}
\ee
in the plaquette partition function, arising after the axial gauge choice.
A relevant open problem is to see if
this deformation of the  GWW model 
can arise in a natural way in the context of
the original 2d lattice gauge theory
(for example, by starting 
with a Hamiltonian of the Kogut-Susskind type).
Clearly, a deeper understanding of a possible lattice gauge theory origin of the model is desirable. This seems relevant considering also that low dimensional lattice gauge theories are being understood nowadays with modern tools from quantum information theory (such as matrix product states or tensor networks) and from the point of view of quantum simulation \cite{Banuls:2019bmf}.

Alternatively, one could view the $\tau$ deformed matrix model as a phenomenological
model which may incorporate some interesting physical features of gauge theory.
The most dramatic effect of the deformation is to smooth out the GWW phase transition.
For any small positive $\tau $, the transition becomes a cross-over.
As a result, there is no ungapped phase where eigenvalues get distributed 
over the whole circle in the ``physical" region $\tau>0$.
In addition, there are many new features in other regions of parameter space, including various phase transitions, which, perhaps, could eventually find some interesting physical interpretation in the context of gauge theory.

In particular, in the $\lambda <0$ region the deformation has a different effect.
We recall that one can switch the sign of $\lambda $ by a formal shift in the angular variables and view the model as the usual GWW model with $\lambda>0$
deformed by a $2\tau \ln |\sin\frac{\varphi}{2}|$ term in the potential. This corresponds to the insertion of an operator
\be
\tilde O^\nu\ ,\qquad \tilde O\equiv \frac{1}{4}\det(1-U)\det(1-U^\dagger)\ .
\ee
With this deformation, the GWW transition extends to a critical line 
$|\lambda_{\rm cr}(\tau)|$, given in \eqref{lamcri}, and in the new phase eigenvalues
 get distributed into two 
separated, symmetric cuts. This is the counterpart of the third order phase
transition found in the dual Hermitian model \cite{Russo:2020pnv}.




An interesting question is if all phases appearing in the whole $(\lambda, \tau)$
parameter space have a counterpart in the Hermitian model of \cite{Russo:2020pnv}. 
The correspondence  in the phase structure of unitary and Hermitian matrix models in the large $N$  limit  involves subtle issues that have been recently discussed in \cite{Santilli:2020ueh}. 
In the present model, the main difference will occur in the $\tau<0$ region. In the
unitary model there are normalizable solutions with eigenvalues sitting at the wells $\pm \pi$; in the Hermitian model eigenvalues will be spread to infinity and the corresponding solution would not be normalizable.
In some unitary models the underlying mechanism triggering some phase transitions
may be different. See for example a discussion in Appendix (B.1) in \cite{Santilli:2018byi}.

It would also be very interesting to understand the contribution of large $N$ instantons by
computing complex saddle-points. This analysis was carried out only recently for the GWW model \cite{Buividovich:2015oju,Alvarez:2016rmo} (see also \cite{Okuyama:2017pil}). 
The GWW model describes certain quantum amplitudes of the XX spin chain model \cite{Perez-Garcia:2013lba}. If $B=\I t$ with $t$ real, the amplitude is related to real-time dynamics of the spin chain model, but then there is no phase transition \cite{spinchain}. In general, it is relevant, in that context, to study the case of a GWW model with a complex parameter \cite{spinchain,Santilli:2018byi}. It would then be interesting to study the eventual phase structure for imaginary or even complex values of at least one of the parameters of the model.


In addition to lattice gauge theory and spin chain models, there are a number
of extremely insightful applications of the GWW model in  low-energy QCD \cite{Leutwyler:1992yt,Verbaarschot:2005rj}. In that context, insertions of the characteristic polynomial type, such as \eqref{insertion}, appear often. In the specific case of the GWW model,  a simpler insertion of the form $\det(U)^{\nu}$ (which has an interpretation as a topological term) has been considered in \cite{Rossi:1996hs,Leutwyler:1992yt,Verbaarschot:2005rj},

\subsection*{Acknowledgements}

J.G.R. acknowledges financial support from projects 2017-SGR-929, MINECO grant FPA2016-76005-C. The work of MT was partially supported by the Funda\c{c}\~{a}o para a Ci\^{e}ncia e a Tecnologia (FCT) through FCT Project PTDC/MAT-PUR/30234/2017.

\end{document}